\documentclass[prl,amsmath,amssymb]{revtex4-1}

\usepackage{subcaption}
\usepackage{graphicx}
\usepackage{newtxtext}
\usepackage{newtxmath}
\usepackage{natbib}
\usepackage{hyperref}
\hypersetup{
    colorlinks = true,
    urlcolor   = blue,
    citecolor  = black,
}

\newcommand{\RomanNumeralCaps}[1]
\linenumbers

\begin{document}

\title{Birth of a bubble: Drop impact onto a thin liquid film for an immiscible three-fluid system}

\author{Pierre-Antoine Ma\"es} 
\affiliation{LadHyX, CNRS \& Ecole Polytechnique, UMR 7646, IP Paris, 91128, Palaiseau, France}

\author{Alidad Amirfazli}
\affiliation{Department of Mechanical Engineering, York University, Toronto, ON M3J 1P3, Canada.}

\author{Christophe Josserand}
\affiliation{LadHyX, CNRS \& Ecole Polytechnique, UMR 7646, IP Paris, 91128, Palaiseau, France}

\begin{abstract}
When a drop impacts a solid substrate or a thin liquid film, a thin gas disc is entrapped due to surface tension, the gas disc retracts into one or several bubbles. While the evolution of the gas disc for impact on solid substrate or film of the same fluid as the drop have been largely studied, little is known on how it varies when the liquid of the film is different that of the drop. We study numerically the latter unexplored area, focussing on the contact between the drop and the film, leading to the formation of the air bubble. The volume of fluid method was adapted to three fluids in the framework of Basilisk solver. The numerical simulations show that the deformation of the liquid film due to the air cushioning plays a crucial role in the bubble entrapment. A new model for the contact time and the entrapment geometry was deduced from the case of the impact on a solid substrate. This was done by considering the deformation of the thin immiscible liquid layer during impact depending mainly on its thickness and viscosity. The lubrication of the gas layer was found to be the major effect governing the bubble entrapment. However the film viscosity was also identified as having a critical role in bubble formation and evolution; the magnitude of its influence was also quantified.
\end{abstract}

\maketitle

\section{Introduction}
Droplet impacts are essential in very different contexts from industrial applications such as ink-jet printing, additive manufacturing or atomization of liquid jet to environmental situations such as soil erosion by rain. The drops can impact a solid surface~\citep{josserand_drop_2016}, a thin liquid film~\citep{josserand_droplet_2003,yarin_drop_2006}, or a deep liquid pool~\citep{duchemin_jet_2002,michon_jet_2017}. \\
Although the dynamics of the impacts depends strongly on the impacted substrate, all cases exhibit similar general trends, for instance, spreading at low impact velocity and splashing at high impact velocity. Another feature seen in drop impact is the entrapment of an air pocket at the moment when the drop touches the substrate~\citep{thoroddsen_air_2003,thoroddsen_air_2005}, giving birth to a bubble. This bubble entrapment is due to an increase of the pressure in the thin air layer located between the drop and the surface prior to impact because of viscous cushioning~\citep{korobkin_trapping_2008,mandre_precursors_2009,hicks_air_2011,duchemin_curvature_2010}. It has been characterised both experimentally \citep{thoroddsen_air_2005,veen_direct_2012,thoraval_drop_2013}, theoretically, and numerically \citep{hicks_air_2011,mandre_precursors_2009,duchemin_rarefied_2012,jian_split_2020} for impact onto solid surfaces or when drop impacts the surface of the same liquid. 
Because the air layer cushioning follows a lubrication dynamics, the bubble geometry can be simply described by the Stokes number, defined here using the drop diameter $2R_0$ and as the ratio between the gas viscous force with the drop inertia $St=\frac{\mu_{g}}{2\rho_{l}U_{0}{R_0}}$,  $\rho_l$ being the drop liquid density, $U_0$ its impacting velocity, and $\mu_{g}$ the viscosity of the gas~\citep{korobkin_trapping_2008,mandre_precursors_2009,duchemin_rarefied_2012,klaseboer_universal_2014}. Then balancing the drop inertia with the gas lubrication in the short time of impact, one can obtain scaling laws for the bubble entrapment \citep{korobkin_trapping_2008,mandre_precursors_2009,bouwhuis_maximal_2012,duchemin_curvature_2010}: the dimensionless impact time and the thickness of the entrapped gas disc scale with $St^{2/3}$, while the dimensionless radius is following a $St^{1/3}$ law.\\
After its formation, the entrapped bubble usually reaches the free surface where it breaks, ejecting small micro-droplets. Understanding the bubble entrapment mechanism and its rupture is thus crucial, for instance, for printing in order to control side ejections or in aeronautics for de-icing film protection~\citep{yamazaki_review_2021}.\\
In this latter case, a droplet of water impacts a film of another liquid (typically glycol) and the two liquids are usually immiscible. For when bubble entrapment involves three fluids, its formation is related to the formation of a triple point between these three fluids.
While bubble formation with two fluids (air and drop on solid or same liquid substrate) has been largely studied, bubble formation and dynamics when three fluids (air, drop and film liquid) are involved  remains largely unexplored, despite its importance in many different situations, from emulsion in food industry to geophysical flows~\citep{yeganehdoust_numerical_2020,alventosa_inertio-capillary_2022}. When drop impact is considered and the two fluids are immiscible, the transition between rebound and coalescence differs \citep{pan_controlling_2016,yeganehdoust_numerical_2020,huang_transitions_2021}, while one can also question the general features of the splash formation \citep{marcotte_ejecta_2019}. The aim of this paper is thus to characterize numerically how the bubble entrapment (dynamics and geometry) is influenced by the presence of a thin film of dissimilar liquid. We take advantage of the Basilisk free software \citep{popinet_accurate_2009} that we adapted to model three immiscible (denoted later with air, water, and oil) fluids using three volume of fluid functions. \\
In fact, three fluids numerical simulations have been the subjects of a few recent studies, e.g., drop impacts on deep pools \citep{hendrix_universal_2016,fudge_dipping_2021,sykes_droplet_2023}, and phase change and atomization \citep{dritselis_open-source_2022,mcginn_unified_2022,yakush_three-phase_2024} in particular. On the numerical method itself, the main challenge lies in the determination of the correct surface tension terms \citep{zhao_hybrid_2024,kromer_efficient_2023}. This issue has been dealt with different platform differently, for instance OpenFoam \citep{dritselis_open-source_2022,mcginn_unified_2022,yakush_three-phase_2024} or using level-set modelling \citep{zhao_finite_2016,yap_numerical_2017}. Nevertheless to the best of our knowledge, we are aware only of a few studies of drop impact on a thin film of another liquid \citep{chen_drop_2017,che_impact_2018,qin_effects_2023,sanjay_drop_2023,fudge_drop_2024}. Finally, in the present paper, we will focus on the birth of the bubble, therefore, just prior the formation of the triple point (so that the no triple line dynamics has to be considered) and propose a model for the bubble entrapment in this configuration.

\section{General problem statement and numerical method}

\begin{figure}
    \centerline{
    \includegraphics[]{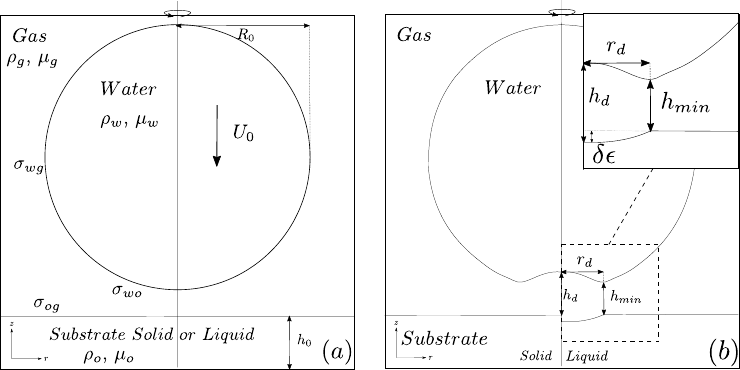}}
    \caption{Left: impact geometry with the different parameters of interest. Right: graphical description of the deformed droplet prior to the contact, illustrating the different measured quantities of interest, the thickness of the gas layer under the center of the droplet $h_{d}$, the minimal distance between the droplet and the substrate $h_{min}$ and the radius of the dimple $r_{d}$. Right and left part of the figure indicate the difference between the impact on solid or liquid substrate with the introduction of $\delta \epsilon$ to describe he deformation of the oil film.}
     \label{fig:variables}
\end{figure}

As shown on Figure \ref{fig:variables} a), a drop of diameter, $D_0=2R_0$, typically made of water, characterized by $\rho_w$ and $\mu_w$ as its density and dynamic viscosity, respectively, impacts a solid substrate or a thin film of thickness $h_0$, at velocity $U_0$; the oil film is characterized similarly by $\rho_o$ and $\mu_o$. The density and dynamic viscosity of the surrounding air are denoted $\rho_g$ and $\mu_g$. The surface tensions associated with the different interfaces are $\sigma_{wg}$, $\sigma_{wo}$ and $\sigma_{og}$, (subscripts, o, g, and w refer to  oil, gas, and water interfaces) for impact on a solid substrate, it simplifies onto one surface tension $\sigma_{wg}$ and the (static) contact angle $\theta$ through application of Young equation.

\subsection{Physical variables and dimensional analysis}

Drop impact onto a surface depends {\it a priori} on many different parameters (drop diameter and velocity, the properties of the fluids and their interface, the oil film thickness, and the gravity), leading to numerous dimensionless numbers.
The two important dimensionless numbers that describe the dynamic features of the impact (therefore based on the drop velocity), namely are the Reynolds and the Weber number (based on the liquid drop and the gas/liquid interface), which balance the inertia with the viscous and capillary effects, respectively, and given as:
$$ {\rm Re}= \frac{2\rho_w R_0 U_0}{\mu_w} \;\; {\rm and} \;\; {\rm We}= \frac{2\rho_w R_0 U_0^2}{\sigma_{wg}}.$$
Alternatively, one can use the Ohnsesorge number which does not involve the velocity, to one of the above numbers:
$$ {\rm Oh}= \frac{\mu_w}{\sqrt{2\rho_w R_0 \sigma_{wg}}}.$$
The gravity effect can be quantified using the Bond number:
$$ {\rm Bo}=\frac{2gR_0}{U_0^2},$$
For millimetric droplets impacting at meter per second speed, the Bond number remains of the order or below $10^{-2}$, indicating that the gravity can be neglected during the impact. Therefore, although it is important to accelerate the drop before the impact, gravity will be neglected for the impact process analysis.\\
In the present study we will assume that the dynamics in incompressible. In fact, while the Mach numbers in both liquid are small so that compressibility can be safely neglected in the liquids, the situation is more complex in the lubrication gas layer, as investigated by \citep{mandre_precursors_2009}. There, they have shown that compressible effects become relevant when the dimensionless number $\epsilon=\frac{P_0}{(\rho_w^4R_0U_0^7\mu_g^{-1})^{1/3}}$, where $P_0$ is the ambient pressure, is small ($\epsilon \ll 1$). For the typical values of the physical parameters used in our paper, $\epsilon \sim 5$ for $U_0= 1 {\rm m}\cdot {\rm s}^{-1}$ and goes down to $0.12$ for the highest velocities studied. Notice that we are using a more viscous (10 times) gas than the air. It does not change qualitatively our results but allows for more reasonable time calculations.
There, the compressibility lowers the height of the bubble entrapment by a factor 2 at most. Therefore, even if compressibility might affect our results for the highest velocities investigated, it can be safely neglected for most of our study. Furthermore, the goal of our paper is to investigate and quantify the effect of the oil layer for incompressible dynamics, and the additional effect of the compressibility should be the subject of future studies.
Depending whether the impact is on a solid or an oil film, several additional dimensionless numbers can be obtained by considering ratios between the different fluids physical properties or geometrical quantities, such as:
\begin{equation}
    \theta, \;\; \frac{h_0}{R_0}, \;\; \frac{\mu_g}{\mu_w}, \;\; \frac{\mu_o}{\mu_w}, \;\; \frac{\rho_g}{\rho_w}, \;\; \frac{\rho_o}{\rho_w}, \;\; \frac{\sigma_{wo}}{\sigma_{wg}} \;\;
{\rm and} \;\; \frac{\sigma_{og}}{\sigma_{wg}}.
\label{eq : param}
\end{equation}

Whereas the parameters in eq \ref{eq : param} may have some influence on the dynamics of drop impact, a systematic study varying all these parameters is out of the scope of this paper. Instead, we will focus on the difference between impacts on a solid substrate with impacts on thin oil films (thus for small ratio $\frac{h_0}{D_0}$), varying primarily the impact velocity and the oil viscosity. It is important to emphasize that for impacts on solid substrates or thin films of the same fluid, it has been
shown that the dynamics at short time (the one of interest for us) was controlled mostly by the air cushioning through the Stokes number defined by:

\begin{equation}
    St=\frac{\mu_{g}}{2\rho_{w}U_0R_0},
    \label{St nb}
\end{equation}
with a small dependence on the surface tension~\citep{klaseboer_universal_2014}.
The aim of the present paper is thus to investigate and characterize how the oil film influence this short time dynamics through the bubble entrapment.

\subsection{Numerical method}

An innovative three fluids volume of fluid (VOF) method was developed in the framework of the \textit{Basilisk} software~\citep{popinet_accurate_2009}. The one fluid formulation of the incompressible Navier-Stokes equation was adapted to describe
three fluids system~\citep{tryggvason_direct_2011}:

\begin{equation}
    \begin{cases}
    \rho^{*}\left (\frac{\partial \Vec{u}}{\partial t}+(\Vec{u}.\nabla )\Vec{u} \right )=-\nabla P + \nabla.(\mu^{*}\nabla\Vec{u})+\sigma^{*} \kappa \delta_{s} \Vec{n},\\
    \nabla.\vec{u}=0\\
    \end{cases}
    \label{eq1fluide}
\end{equation} 
where $\rho^{*}$ and $\mu^{*}$ are constant domain functions accounting for the values of the densities and viscosities in each fluid. The Dirac function $\delta_{s}$, the curvature $\kappa$ and the normal $ \Vec{n}$ are defined on the interfaces, while $\sigma^{*}$ is also a constant by domain function adapted for the interfaces. The parameters $\rho^{*}$, $\mu^{*}$ and $\sigma^{*}$ are defined using three characteristic functions $\chi_w$, $\chi_g$ and $\chi_o$ whose value is one in the associated domains ($w$, $g$ and $o$) and zero otherwise, such that one can define:
\begin{equation}
    \rho^{*}=\rho_w \chi_w+\rho_g \chi_g+\rho_o \chi_o \,\, {\rm and} \,\,  \mu^*=\mu_w \chi_w+\mu_g \chi_g+\mu_o \chi_o.
    \label{char}
\end{equation}
It is important to note that in this approach, each physical interface (say $wg$) corresponds to the jump of two characteristic functions ($\chi_w$ and $\chi_g$). A surface tension $\sigma_j$ with $j={w,g,o}$ is associated to each characteristic function, which can then be easily related to the physical surface tensions through the relation:

\begin{equation}
    \begin{cases}
    \sigma_{w}=\frac{1}{2}(\sigma_{wo}+\sigma_{wg}-\sigma_{og})\\
    \sigma_{g}=\frac{1}{2}(\sigma_{wg}+\sigma_{og}-\sigma_{wo})\\
    \sigma_{o}=\frac{1}{2}(\sigma_{og}+\sigma_{wo}-\sigma_{wg}).\\
    \end{cases}
\end{equation}

The $\sigma_j$ correspond to the Antonoff's rule~\citep{lang_interfacial_1976}, indicating that if one value of the $\sigma_j$ is negative (in fact only one can be negative for positive surface tension), say $\sigma_w$ for instance, then at equilibrium water will always separate from the two other fluids, since the the oil-gas interface energy is larger that the sum of oil-water and gas-water interfaces. On the other hand, when all $\sigma_j$ are positive, the equilibrium exists and the Neumann triangle condition \citep{david_investigation_2009} prevails at the triple point.
With these definitions, the one fluid formulation equations \ref{eq1fluide} describes now the three-fluids dynamics straightforwardly and its numerical discretization follows that of the one developed for two fluids. In particular, we use the VOF method to solve the set of equations: a colour function ($c_j$ ) is defined and associated to each characteristic function, so that in each computation cell, then we have :

\begin{equation}
    \rho^{*}=\rho_w c_w+\rho_g c_g+\rho_o c_o \,\, {\rm and} \,\,  \mu^*=\mu_w c_w+\mu_g c_g+\mu_o c_o,
    \label{color}
\end{equation}
where the $c_j$ correspond to the volume fraction of each fluid in the cell, imposing the extra condition, for each cell, so that:
\begin{equation}
c_w+c_g+c_o=1.
\label{eq:3c}
\end{equation}
Whereas this condition is always satisfied, by definition, for two fluids (where formally $c_w=c$ and $c_g=1-c$), it is not the case {\it a priori} for three fluids. In fact, nothing ensures that the condition (\ref{eq:3c}) is preserved by the numerical integration of the set of equations.

\begin{figure}
    \centering
    \includegraphics[scale=2]{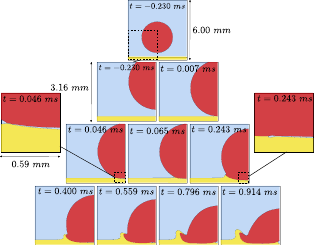}
    \caption{Snapshots of the impact at several time step (indicated on the images), with different geometrical sizes. The top image shows the initial condition, while the pyramidal images below it present the dynamics at scale two. Two zoom images (scale 1/10) on the left and right of the pyramid illustrate the formation and retraction of the air film entrapped by the impact. The first contact between the water droplet (red) and the oil substrate (yellow) occurs around the $46$ $\mu s$. Then a small disc of gas (in blue) is entrapped and retracts, while a corona made of a mix of oil and water emerges from the impact at larger times.}
    \label{fig : pannel drop impact}
\end{figure}

For the short times studied here, we can assume angular symmetry  so that the calculation will be done as a 2D-axisymmetric system. This is reasonable for case ${\rm We}$ numbers studied here are well below splash threshold. The calculation domain is a square box of size $6 mm \times 6mm$. The diameter of the drop is $D_0=2 R_0=3*10^{-3}$ $m$, water $\rho_w=1000$ $kg.m^{-3}$ was used as drop liquid with slightly higher viscosity $\mu_w=0.002$ $Pa.s$ for numerical convenience, with no qualitative changes in the results.\\
Adaptive mesh refinement was used to achieve better resolution: the mesh refinement is controlled by both the velocity gradient and the interface location, down to $1/2^{12}$ (refinement level 12, corresponding to a minimum cell size of $1.46$ $\mu m$). 
The impact velocity was varied between $0.5$ and $5$ $m.s^{-1}$. To avoid numerical instabilities in the gas phase, its density was taken slightly larger than the air one, $\rho_g=2$ $kg.m^{-3}$ and we have checked that it does not influence our results. The gas viscosity was varied from $7.10^{-5}$ to $5.10^{-4}$ $Pa.s$ to achieve specific values of the Stokes number, with a usual value $\mu_g=1.6*10^{-4}$ $Pa.s$ for most of our simulations. The viscosity of the oil film was varied from $0.002$ to $0.7$ $Pa.s$, whereas the thickness $h_0=0.3$ $mm$ , and the density $\rho_o=900$ $kg.m^{-3}$ remained unchanged. The surface tension values for water, air and oil were: $\sigma_{wg}=7.0*10^{-2}$, $\sigma_{wo}=2.0*10^{-2}$ and $\sigma_{wg}=3.2*10^{-2}$ $N.m^{-1}$, given that each $\sigma_j$ is positive so that the triple point exists. The time $t=0$ is defined as the mathematical time at which the drop outline (a circle) falling at velocity $U_0$, would intersect the substrate plane. 
A typical impact is shown in Figure \ref{fig : pannel drop impact}, for $U_0=1.5$ $m.s^{-1}$ and $\mu_o=0.02 \,$ Pa.s to illustrate the general dynamics. Although specific treatment was used for the triple point, we have checked that the dynamics does not vary qualitatively as we vary the resolution. As the drop approaches the oil film, both the drop underneath surface and the oil film deform until a thin layer of air is entrapped when the drop touches the oil film (left zoom image). Later, this entrapped air film retracts because of surface tension until it forms a bubble underneath (see the last image for $t=914 \, \mu$s). On the other hand, the impact generates a corona made of both water and oil at later times.

In this paper, we thus focus on the physical characteristics of the bubble entrapment, e.i., formation time and geometrical dimensions.

\subsection{Definition of the contact}

It is worth noting that since the gas layer obeys the lubrication regime, mathematically the contact between the drop and the film cannot happen in a finite time in the presence of surface tension by only considering the Navier-Stokes equations \citep{duchemin_curvature_2010,yeganehdoust_numerical_2020}. In fact, the physical contact arises rapidly after the gas film thickness becomes locally lower than a microscopic cut-off length scale usually of the order of few hundreds of nanometers below which other mechanisms dominate: surface interactions, non-continuum effects or thermal fluctuations, for instance (see the recent review \citep{sprittles_gas_2024} for a comprehensive analysis of these effects).  Numerically, the contact occurs as soon as a triple line can exist that is when the three phases are present in the same cell. It happens thus when the gas layer reaches the size of the minimal mesh size, which plays here the role of the cut-off length.
Therefore, we define later the contact as when the minimal gas film thickness $h_{min}(t)$, (see Figure \ref{fig:variables}) attains the size of our smallest mesh (here $1/2^{12}$ of our computational domain, corresponding to a cutoff length of $3.7$ $\mu m$). Although our numerical cut-off is at least one order of magnitude larger than the usual microscopic length, we numerically verified that our results were not depending qualitatively on the precise value of this length (see Appendix A). Finally, it is important to note that in the present study, since we determine the bubble entrapment just when a triple point appears, the numerical issues related to the treatment of the triple point does not play a role, neither does the contact angle for impacts onto a solid substrate. Such details are considered within the context of a current comprehensive study.

\section{Results and film effects}

To establish a baseline, first we present the results for birth of the bubble for drop impact onto a solid surface. Dimensionless variables are shown using the overbar notation.
\subsection{Drop Impact over a solid}

\begin{figure}
    \centering
    \includegraphics[scale=0.45]{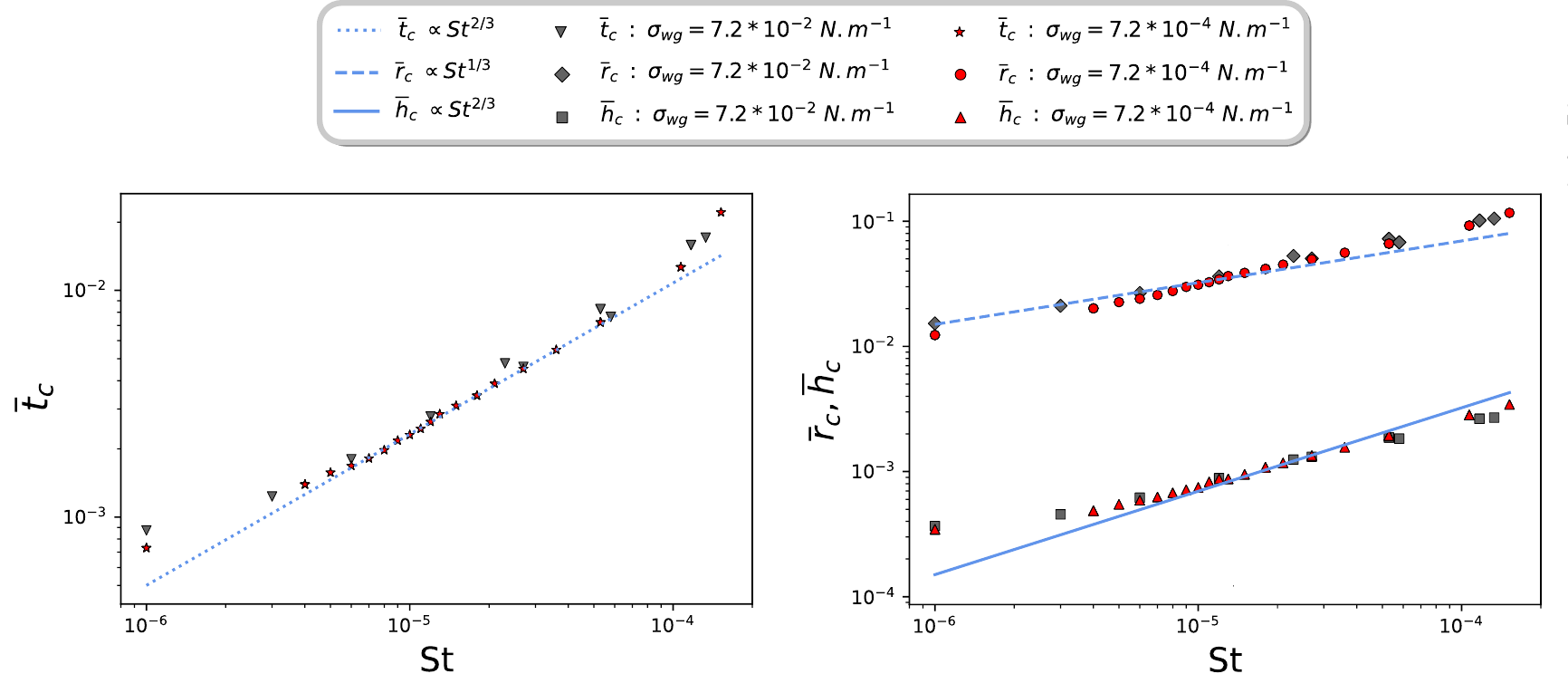}
    \caption{(left) Dimensionless time of contact $\bar{t}_c$ of a drop impacting on a solid as function of the Stokes number $St$. The data set are obtained for two surface tension, grey $\sigma_{wg}=7.2 \cdot 10^{-2}$ and red $\sigma_{wg}=7.2*10^{-4}$ $N \cdot m^{-1}$, varying both the impact velocity and the gas viscosity. The expected scaling law $St^{2/3}$ is indicated (blue line); (right) the dimensionless radius $\bar{r}_c$ (upper curve) and central height $\bar{h}_c$ (lower curve). Both are in agreement with the predicted scaling laws $St^{1/3}$ (resp $St^{2/3}$).}
    \label{fig : t solid impact}
\end{figure}

Figure \ref{fig : t solid impact} left shows the contact time between the impacting drop and the solid substrate as a function of the the Stokes number. The Stokes number is varied by changing the impact velocity, but also by modifying the gas viscosity to cover a wide range of values. A good agreement between the data with the $St^{2/3}$ scaling law predicted by the models balancing the inertia of the impact with the gas lubrication is observed~\citep{mandre_precursors_2009,duchemin_curvature_2010,bouwhuis_maximal_2012}. Results for two different values of the gas-liquid surface tension,  $\sigma_{wg}=0.072 \; N\cdot m^{-1}$ (grey points) and for a surface tension one hundred time smaller (in red), show that the surface tension has very little influence on the contact in the frame of our study, as assumed in the theoretical predictions. Our data set is bound at high Stokes number by the finite size effect: indeed for $St>10^{-3}$ the contact diameter has the same order of size as the droplet diameter so that the hypothesis leading to the scaling law (see model below) is not applicable anymore. On the other hand, at low Stokes numbers (below $10^{-6}$) the limit of the resolution is clearly reached. 
Recall the main physical ingredients of the model that predicts the $St^{2/3}$ scaling \citep{duchemin_curvature_2010} is based on the comparison between the lubrication pressure due to the cushioning of the air, and the pressure needed to deform the droplet when approaching the substrate. If air pressure is larger than that to deform the droplet, the contact with the substrate is avoided, and the contact time is when these two pressures balance. To estimate these pressures, we use the scaling argument that the radial length scale (noted here generically as $r_d(t)$, see Figure \ref{fig:variables} right) of the deformation is related to the vertical scale of the drop deformation ($h_d(t)$), through the geometry of the drop $ r_d =\sqrt{2 R_0 h_d}$ notably highlighted by \citep{philippi_drop_2016}. We can then relate the vertical scale to the time through $h_d=U_0 |t|$. For $t<0$, this vertical scaling corresponds to the air layer between the falling drop and the substrate, while for $t>0$ it indicates the height of the drop that has been deformed horizontally because of the substrate.
The lubrication equation for the gas layer thickness $h(r,t)$ in axisymmetric geometry is :
\begin{equation}
    \frac{dh}{dt}=\frac{1}{r}\frac{\partial}{\partial r}\left (\frac{h^{3}r}{3\mu_g}\frac{\partial P}{\partial r}\right ),
\label{eq:lub}
\end{equation}
where $P$ is the pressure in the gas film.
Using the geometrical relations for the horizontal and vertical scales introduced above, we obtain the following scaling for the lubrication pressure in the gas as a function of time:
\begin{equation}
P_{lub} \sim \frac{\mu_g r_d^2 U_0}{h_d^3} \sim \frac{\mu_g R_0}{U_0 t^2}.
\label{eq:pin}
\end{equation}
This lubrication pressure is responsible for the droplet deformation prior to the contact with the substrate. We thus need first to determine when it is enough to delay this contact. For this purpose, one has to estimate the inertial pressure $P_{in}$ that is needed to deform horizontally the drop when approaching the substrate. We can consider that the deformation is located in the half-sphere of radius $r_d$ inside the droplet,
leading to the scaling for this inertial pressure as :
\begin{equation}
P_{in} \sim \rho_w U_0 \frac{dr_d}{dt}\sim \rho_w U_0^2 \frac{R_0}{r_d} \sim \rho_w U_0^2 \sqrt{\frac{R_0 }{U_0t}}.
\label{eq:pin}
\end{equation}

We can infer, by comparing these two pressures $P_{lub}$ and $P_{in}$, the qualitative behaviour of the effect of the gas layer on the impact process as : for negative times, much before the drop is close to the substrate, the lubrication pressure is much smaller that the inertial one, and the drop falls without deformation. As $|t|$ decreases ($t$ being still negative) the lubrication pressure increases until it becomes larger that the inertial one, at this moment the droplet starts to deform due to the gas cushioning, creating a dimple beneath the drop. The lubrication pressure dominates the inertial one, indicating that the drop can deform enough to avoid touching the substrate even for a positive time! Further on, as the lubrication pressure decreases more rapidly that the inertial one for a positive time (since $P_{lub} \propto \frac{1}{t^2}$ whereas $P_{in} \propto \frac{1}{\sqrt{t}}$ for $t>0$), the gas cushioning cannot support the deformation of the drop any longer and it has to contact the substrate. Bubble formation is thus related to the deformation of the droplet at the contact time a thin air disk is trapped between the substrate and the droplet, the disk will later retract into a bubble due to surface tension. With such dynamics, the contact time can be estimated by balancing the two pressures, leading to:
\begin{equation}
\bar{t}_c=\frac{U_0t_c}{R_0} \sim  St^{2/3}.
\label{eq:tc}
\end{equation}
$\bar{t}_c$ can be interpreted both at the typical time before the impact when the drop starts to deform ($t<0$) and that when the contact with the substrate occurs ($t>0$). Within this framework, the surface tension is neglected, which is validated here by our numerical simulations (see Figure \ref{fig : t solid impact} which showed no variation as the surface tension was reduced by two order of magnitude), although it can slightly affect the dynamics, particularly for the drop deformation~\citep{klaseboer_universal_2014}.
We can also deduce from the time of impact, the scaling for the thickness and the radius of the entrapped bubble at the contact:
\begin{equation}
    \frac{h_c}{R_0}=\bar{h}_c \sim St^{2/3} \;\;{\rm and} \;\; \frac{r_{c}}{R_0}=\bar{r}_{c} \sim St^{1/3}.
\label{eq:hr_dimple}
\end{equation}
Figure \ref{fig : t solid impact} right shows the central height of the drop $\bar{h}_c$ and the dimple radius $\bar{r}_{c}$ at the time of impact as functions of the Stokes number for our numerical simulations of impacts on a solid substrate. Although the best fit analysis of the data would give $0.4\pm 0.05$ for the radius scaling and $0.55\pm 0.05$ for the central height, the agreement with the predicted scaling, shown in Figure \ref{fig : t solid impact}, is still qualitatively good.\\
The bubble formation when a drop impacts a solid is thus mainly controlled by the lubrication of the gas layer. This leads to different scaling, functions of the Stokes number and we now seek how these dynamics is affected when the impact is made on a thin liquid film.

\subsection{Drop Impact over a thin liquid film}

\begin{figure}
\centering
    \begin{subfigure}[t]{0.4\linewidth}
        \includegraphics[scale=0.4]{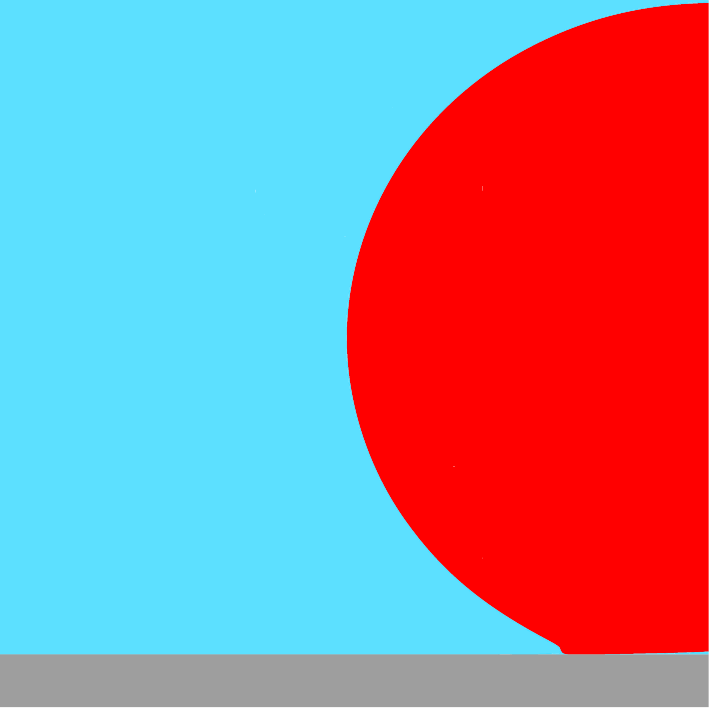}
        \caption{}
    \end{subfigure}
    \begin{subfigure}[t]{0.4\linewidth}
        \includegraphics[scale=0.4]{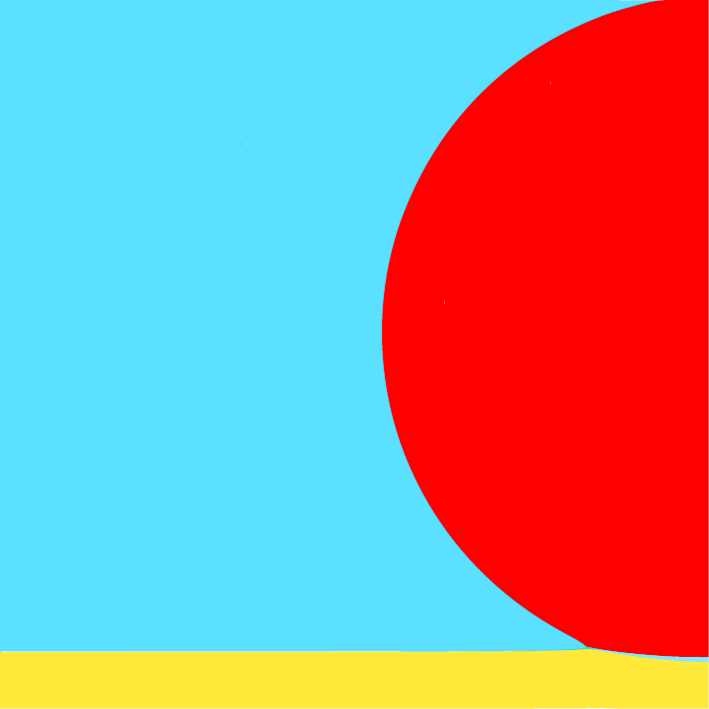}
        \caption{}
    \end{subfigure}
    \caption{Snapshot at $t=0.034$ $ms$ (a) Impact over a solid substrate. (b) Impact on a liquid film. The two initial conditions are exactly the same. $R_0=3$ $mm$, $h_{0}=0.3$ $mm$, $U_0=1$ $m/s$, $\mu_w=0.002$ $Pa.s$, $\mu_o=0.02$ $Pa.s$, $\sigma_{wg}=0.07$, $\sigma_{wo}=0.039$, $\sigma_{og}=0.032$ $N/m$, $Re=1500$, $We_{wg}=43$, $St=5.3*10^{-5}$.}
    \label{fig : compa solid liq}
\end{figure}

When the substrate is replaced by a thin liquid film, the impact differs, as illustrated in Figure \ref{fig : compa solid liq} where a small deformation of the liquid film is also observed.

 Remarkably, the influence of the liquid film on the contact time cannot be deduced {\it a priori}: on one hand, one would expect the contact to happen earlier since the gas cushioning is facilitated by the film due to viscous entrainment \citep{lo_mechanism_2017,duchemin_dimple_2020}; on the other hand, the film deformation should delay the contact. 
\begin{figure}
    \centering
    \includegraphics[scale=0.4]{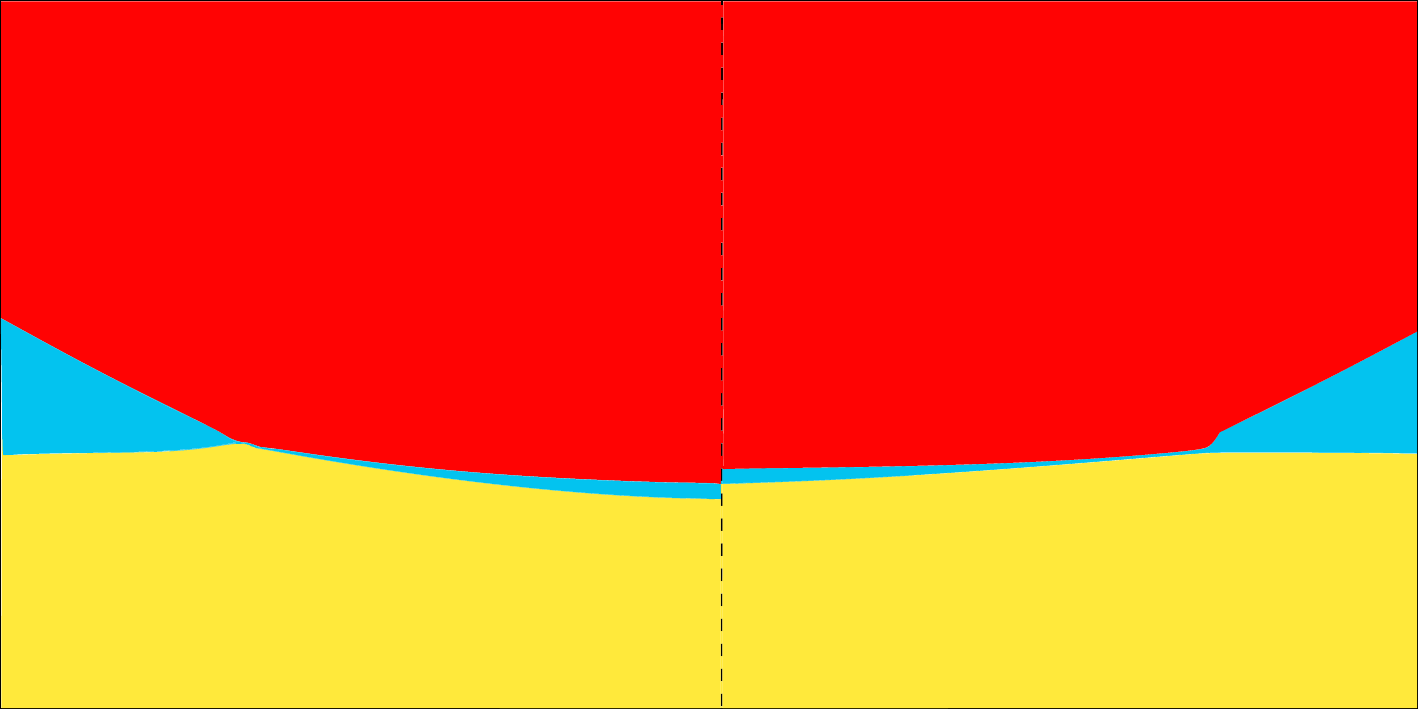}
    \caption{Two snapshots ($t=0.034$ $ms$) for two similar impacts on the different oil (left $\mu_o=\mu_w=0.002$ $Pa.s$, right $\mu_o=0.02$ $Pa.s$). The bubble size is dependent on the oil viscosity. Simulation realized with the following parameters. $St=5.3*10^{-5}$, $U_0=1$ $m/s$, $D_{0}=3$ $mm$, $h_{0}=0.3$ $mm$, $\sigma_{wg}=0.032$, $\sigma_{wo}=0.039$, $\sigma_{og}=0.07$ $N.m^{-1}$.}
    \label{fig : compa 2 visco}
\end{figure}
The film deformation is due to the impact pressure exerted by the drop on the film transmitted by the air layer, as illustrated on Figure \ref{fig:pres-field}. In fact, as explained by \citep{hicks_air_2011}, depending on the impact parameters and the oil film thickness and viscosity, the oil film deformation can be considered inviscid (for low oil viscosity or/and large film thickness) or viscous (thin film and/or low oil viscosity.
The dependence of the deformation with the oil viscosity can be indeed observed in Figure \ref{fig : compa 2 visco}, where a zoom of the entrapped bubble is presented at the contact time for two different oil viscosities $\mu_o=\mu_w=0.002$ $Pa.s$ on the left, and $\mu_o=0.02$ $Pa.s$ on the right, showing that the smaller the viscosity, the larger is the deformation.

\begin{figure}
    \centering
    \includegraphics[scale=0.32]{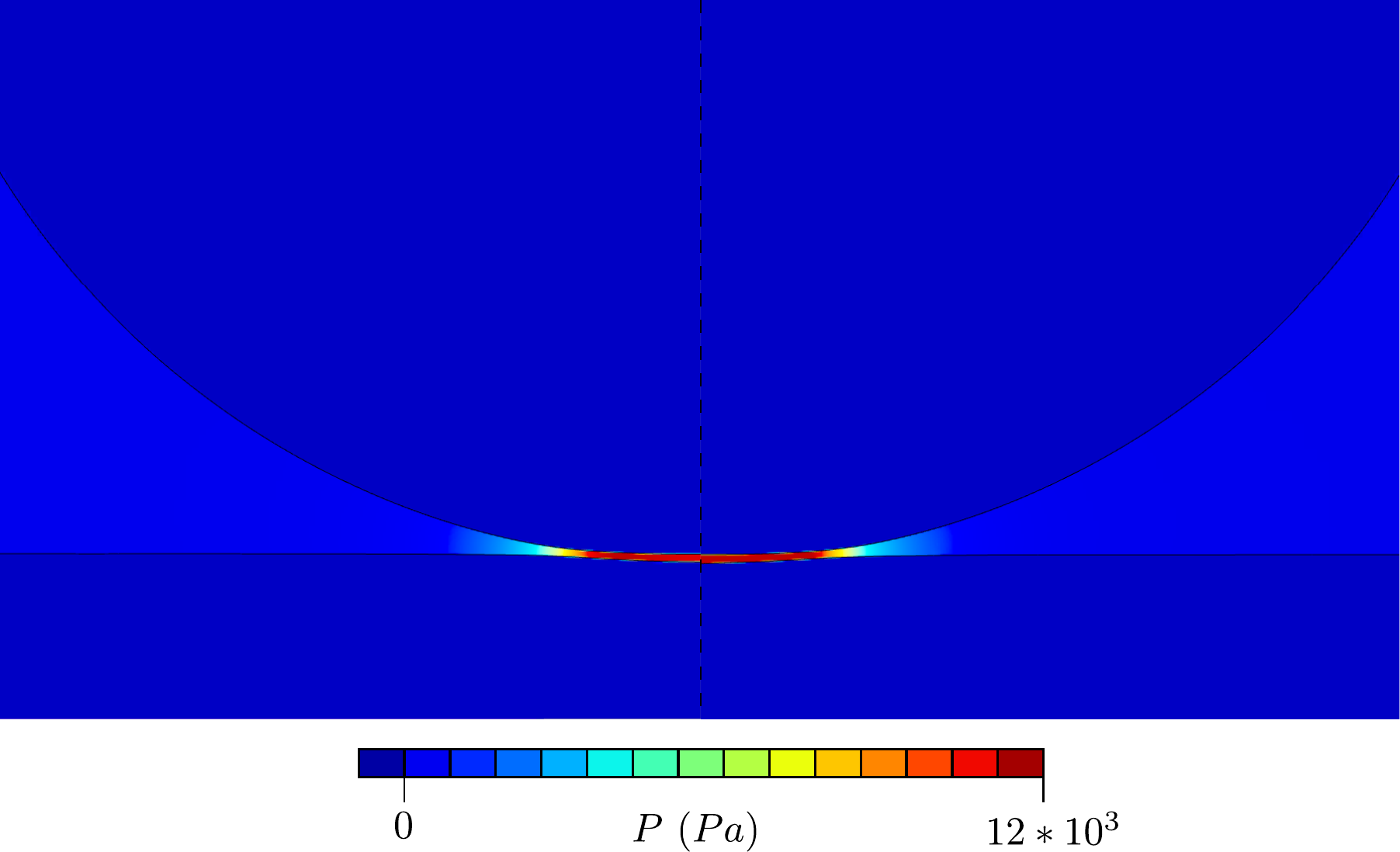}
    \caption{Snapshots of the dimensionless pressure ($t=0.011$ $ms$) for two similar impacts on the different oil (left $\mu_o=\mu_w=0.002$ $Pa.s$, right $\mu_o=0.02$ $Pa.s$). Simulation realized with the following parameters. $St=2.66*10^{-5}$, $U_0=2$ $m/s$, the others parameters being the same as the ones in figure \ref{fig : compa 2 visco}.}
    \label{fig:pres-field}
\end{figure}

\begin{figure}
    \centering
    \includegraphics[scale=0.5]{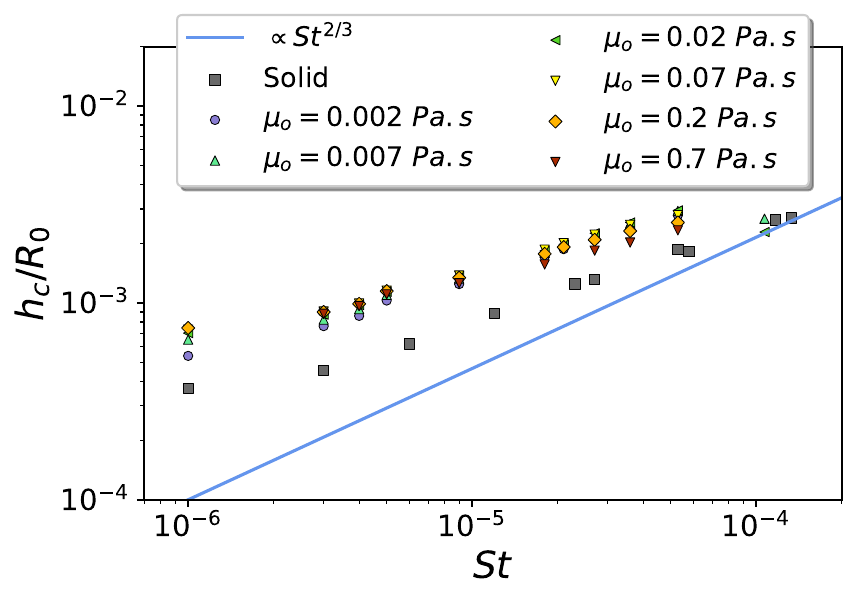}
    \caption{Data set for trend at different Stokes numbers with various oil layer viscosity. The $St^{2/3}$ is still observed. All of these impact cases leads to larger gas layer thicknesses at the contact time compare to the solid surface impact. We also denote a complex dependency in oil viscosity. }
    \label{fig : data all visc}
\end{figure}

To investigate quantitatively the influence of the thin liquid film in the bubble formation mechanism, different sets of numerical simulations were done by varying the impact velocity and different oil viscosities. The results are shown in Figure \ref{fig : data all visc} for thickness $\bar{h}_c$ of the gas layer entrapped at the contact time as function of the Stokes number. One can observe that the gas layer (and the contact time) is systematically higher than that of the impact on the solid substrate, indicating that the delay due to the oil film cushioning dominates as the gas cushioning increase. Furthermore, it can be noticed that these two counteracting effects (Oil cushioning / interface deformation) are not similar for large Stokes numbers (roughly above $10^{-5}$) where the more viscous cases converge towards the solid case. By contrast, for small Stokes number (below $10^{-5}$) the results arrange differently with the oil viscosity.
Finally, it is worth noting that the $St^{2/3}$ scaling still appears to hold qualitatively for the variation of $h_c/R_0$ as function of the Stokes for each oil viscosity. However, performing a best fit analysis for constant oil viscosity data gives lower exponents (around $0.5$) with higher error bar. Therefore, we observe a clear dependence with the oil viscosity that needs to be elucidated.

\subsection{The new scaling law for birth of a bubble}
To account for the influence of the film deformation on the dynamics of bubble formation we first characterize numerically the deformation of the film $\delta \epsilon$ as the difference between the height of the oil surface under the center of the droplet at the contact time and the initial thickness of the oil film (see the left insert in Figure \ref{fig:variables}). Since the oil layer can be deformed, the droplet can continue its fall under the original height of the oil layer before making the contact, which was not possible for a rigid solid. Then it is tempting to adapt the model derived for solid substrate to that with liquid film by taking into account that the drop can now go below the initial surface of the oil film before contacting the oil. This means that the inertial pressure $P_{in}$  needed to deform the drop is lowered because of the film deformation, indeed the substrate is also moving during the fall, and then delays the contact compare to the impact onto solid surfaces. The inertial pressure can thus be deduced as a first approximation from eq. \ref{eq:pin} :
\begin{equation}
    P_{in}=\rho_wU_0^{2}\sqrt{\frac{R_0}{h_d+\delta \epsilon}}.
\end{equation}

On the other hand, if we assume that the scaling for the lubrication pressure eq. \ref{eq:lub} remains unchanged, we obtain the following relations using, as for the drop impact on the solid substrate, the balance between the two pressures:

\begin{equation}
    \frac{h_c}{R_0}\sim St^{2/3}\left(1+\frac{\delta \epsilon}{h_c}\right)^{1/3}, \quad  \frac{r_c}{R_0}\sim St^{1/3}\left(1+\frac{\delta \epsilon}{h_c}\right)^{1/6}, \quad 
    \frac{U_0 t_c}{R_0} \sim St^{2/3}\left(1+\frac{\delta \epsilon}{h_c}\right)^{1/3}.
    \label{eq new scal}
\end{equation}
These relations suggest that the Stokes number of the impact is simply replaced by an effective Stokes number:
\begin{equation}
    St_{eff}=\sqrt{\left(1+\frac{\delta \epsilon}{h_c}\right)}St.
\end{equation}
which involves the oil film deformation $\delta \epsilon$. Remarkably, our argument on the deviation of the pressure because the oil film has deformed is somehow similar that the one used by \citep{howland_its_2016} to explain the delay of the splashing threshold for impact on soft surfaces.

\begin{figure}
    \centering
    \includegraphics[width=0.45\linewidth]{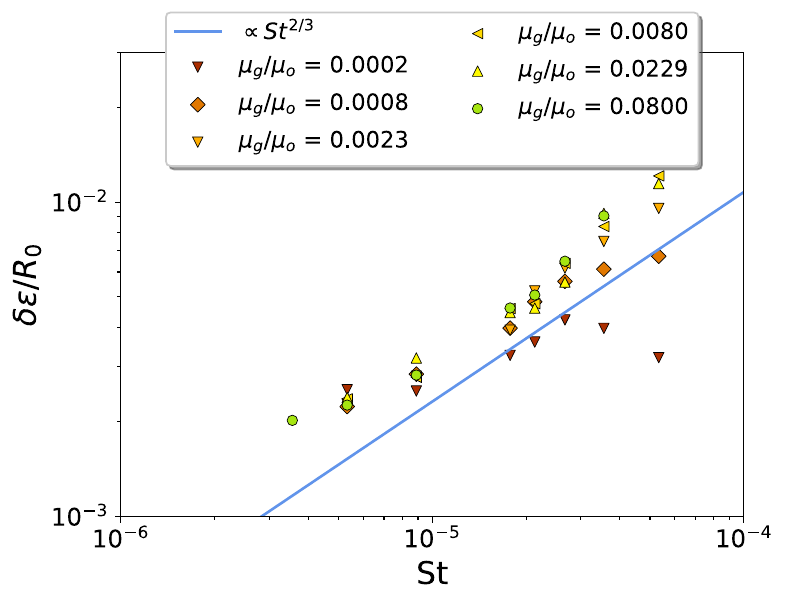} \includegraphics[width=0.45\linewidth]{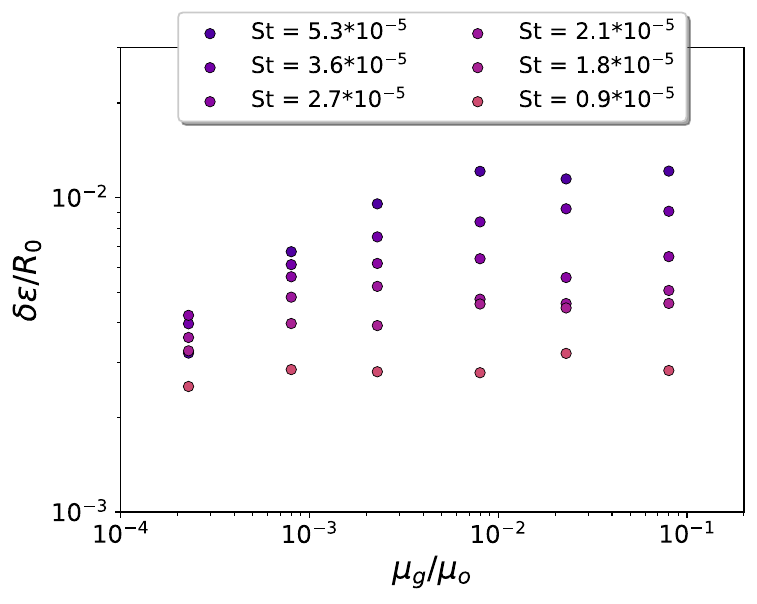}
    \caption{ Deformation of the oil film $\bar{\delta \epsilon}=\delta \epsilon/R_0$ (left) for different viscosity ratios as function of  the Stokes number and (right) for different Stokes number as function of  the viscosity ratio. The line on the left plot indicates the scaling law $St^{2/3}$ consistent with the behaviour of the gas thickness below the droplet. }
    \label{fig : delta eps}
\end{figure}

We just need therefore to determine how $\delta \epsilon$ varies with the impact parameters. Figure (\ref{fig : delta eps}) shows the variation of $\bar{\delta \epsilon}=\delta \epsilon/R_0$ as a function of the Stokes number for different viscosity ratios $\mu_g/\mu_o$ (left), or equivalently as a function of the viscosity ratio for different Stokes number (right). 
We observe that the deformation is the lowest for the highest oil viscosity (low viscosity ratio $\mu_g/\mu_o$) and that it does almost not depend on the viscosity ratio for small Stokes number.

To explain the behaviour of $\bar{\delta \epsilon}$ it is interesting to estimate whether the oil film can be considered inviscid or in the lubrication regime. \citep{hicks_air_2011} have shown that it depends on the ratio between the film thickness $h$ and the viscous length $\left(\frac{\mu_o}{\mu_g}St\right)^{1/3}R$. Therefore, if $\left (\frac{h}{R}\right )^{3} \gg \frac{\mu_o}{\mu_g}St$, the oil film deformation can be considered inviscid either because the oil viscosity is small or the film thickness is large: in both cases the oil layer can deform easily. On the other hand, when the oil viscosity is large or the film thickness small, following $\left(\frac{h}{R}\right)^{3} \ll \left(\frac{\mu_o}{\mu_g}St\right)$, the oil film deforms less easily and obeys a lubrication dynamics. In our study where the drop radius, film thickness are taken constant, the following asymptotic regimes for $\bar{\delta \epsilon}$ are finally observed:

\begin{enumerate}
    \item for small viscosity ratios, corresponding to very viscous oil $\frac{\mu_g}{\mu_o} \ll St \left(\frac{R}{h}\right )^{3}$, we see a small deformation of the oil film. As such, we can deduce the $\bar{\delta \epsilon}$ by using the lubrication equation (\ref{eq:lub}) for the oil film (taking thus $\mu_o$ and a film thickness almost constant $h_o$), the thickness variation using the scaling for the pressure from the impact on solid substrate, yielding the scaling:
    $$\bar{\delta \epsilon} \sim \frac{\mu_g}{\mu_o} St^{-4/3}.$$
    \item For high viscosity ratio $\frac{\mu_g}{\mu_o} \gg St \left(\frac{R}{h}\right )^{3}$, the oil film can deform easily. In this case, the lubrication of the gas film separating the droplet and the liquid film is similar  as eq. (\ref{eq:lub}), the only change being in the prefactor ($1/12$ instead of $1/3$), so that we can deduce: \\
    $$\bar{\delta \epsilon} \sim St^{2/3}.$$
\end{enumerate}

\begin{figure}
    \centering
    \includegraphics[width=0.45\linewidth]{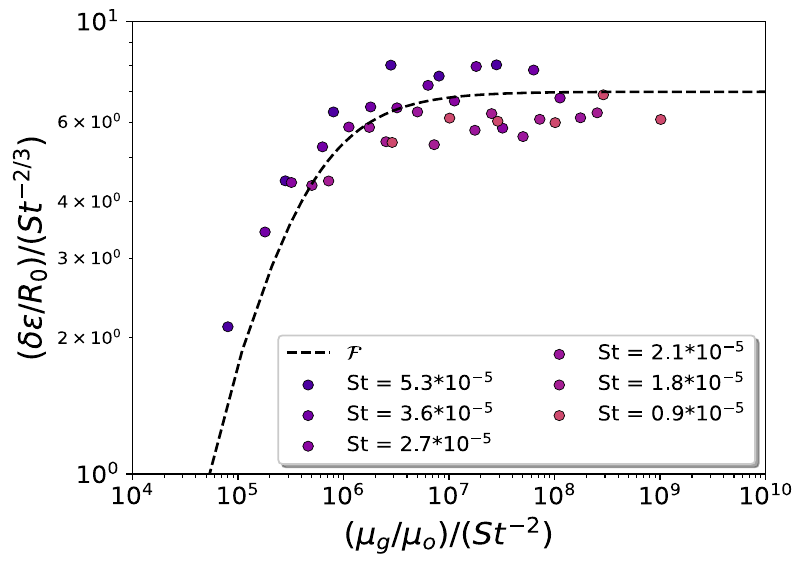} \includegraphics[width=0.45\linewidth]{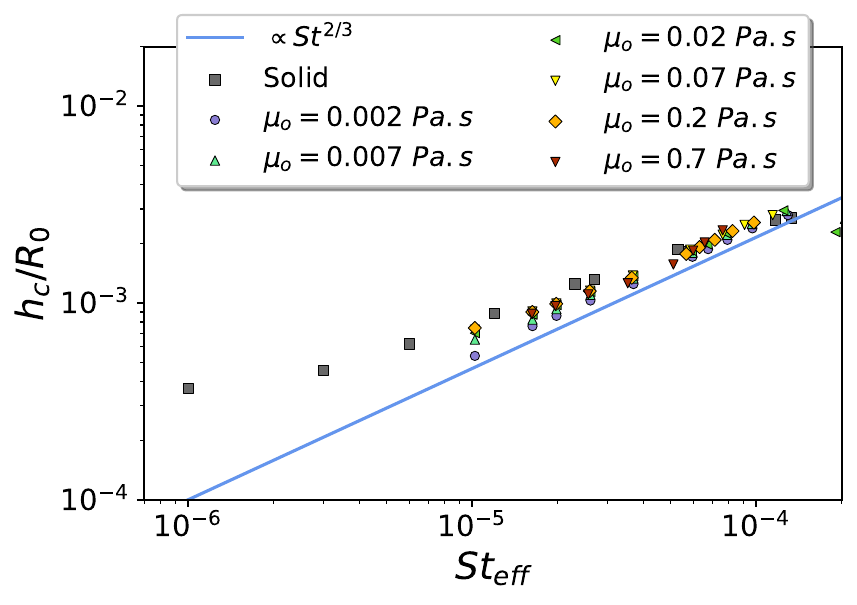}
    \caption{(left) Oil film deformation $\bar{\delta \epsilon}$ divided by $St^{2/3}$, as function of the rescaled variable $\xi=\frac{\mu_g}{\mu_o} St^{-2}$. The Padé approximation accounting for the two asymptotic behaviour is shown as dashed line.  (right) Thickness of the gas layer rescaled  using the new scaling law. We see collapse of the data on the solid line, in agreement with the $2/3$ scaling.}
    \label{fig : delta eps resc}
\end{figure}

These two asymptotic behaviours can be tested by plotting $\bar{\delta \epsilon}St^{-2/3}$ as function of the rescaled variable $\xi=\frac{\mu_g}{\mu_o} St^{-2}$; a linear behaviour for small $\xi$ and a constant one for large $\xi$, is seen in Figure (\ref{fig : delta eps resc}).
These two behaviours can be combined using a Padé approximation, leading to the following expression for the deformation:

\begin{equation}
    \frac{\delta \epsilon}{R_0}St^{-2/3}=A\frac{\frac{\mu_g}{\mu_o}St^{-2}}{B+\frac{\mu_g}{\mu_o}St^{-2}}
    \label{eq scaling delta eps}
\end{equation}

where the constant $A=7$ and $B=3 \times 10^{5}$ can be fitted from our data, as shown on Figure \ref{fig : delta eps resc} (left). These constants still contain all the physics of the parameter which are maintained constant in the frame of our study. 
Using a Padé approximation is a natural way to combine two asymptotic regimes without introducing additional contribution. Beside its fitting features, it can be taken as the simplest approach to associate two regimes only.

Finally, we can deduce from this former scaling on $\delta \epsilon/R_0$ a scaling for $\delta \epsilon/h_c$, using the asymptotic behaviour of $h_c$ for solid substrates ($h_c \sim St^{2/3}$), yielding:

\begin{equation}
    \frac{\delta \epsilon}{h_c}=C\frac{\frac{\mu_g}{\mu_o}St^{-4/3}}{B+\frac{\mu_g}{\mu_o}St^{-2}}
    \label{scaling-eps-hc},
\end{equation}
where $C=7 \times 10^{-3}$ is directly deduce from $A$.
We can now combine the two equations \ref{eq new scal} and \ref{scaling-eps-hc} to obtain the new expressions of the scaling laws for the bubble entrapment:
\begin{eqnarray}
    & \bar{t_{c}}=\frac{U_0 t_c}{R_0}\sim\frac{h_c}{R_0}\sim St^{2/3}\left(1+C\frac{\frac{\mu_g}{\mu_o}St^{-2}}{B+\frac{\mu_g}{\mu_o}St^{-2}}\right)^{1/3}, \\ 
    & \frac{r_c}{R_0}\sim St^{1/3}\left(1+C\frac{\frac{\mu_g}{\mu_o}St^{-2}}{B+\frac{\mu_g}{\mu_o}St^{-2}}\right)^{1/6}
\end{eqnarray}

Note that these scalings are consistent with the expressions deduced for the impact on a solid substrate: indeed, taking a solid as the limit when the oil viscosity diverges, the viscosity ratio goes to $0$ and the previous scalings eq \ref{eq:hr_dimple} are recovered. Using this new scaling we obtain a collapse of all our data for impacts over a liquid film with those of the impact on a solid substrate (Fig \ref{fig : delta eps resc} (right)). 
It therefore highlights and quantifies the influence of the oil viscosity on the entrapment, as such we have shown that the cushioning of the gas layer dominates the bubble entrapment dynamic, but also that the oil film deformation cannot be neglected since it controls the delay of the entrapment. 

\subsection{Other results and discussion}

\begin{figure}
    \centering
    \includegraphics[width=0.45\linewidth]{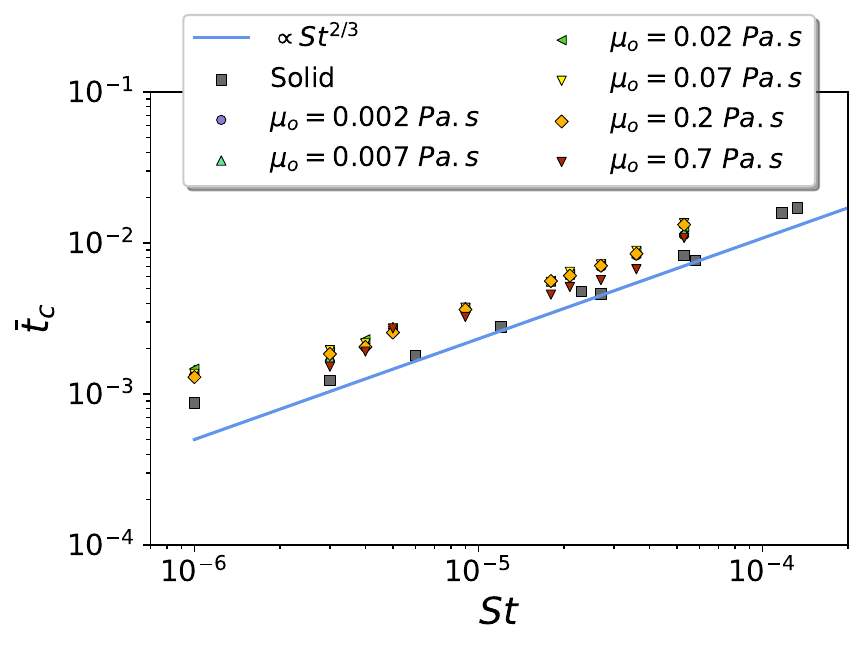}
    \includegraphics[width=0.45\linewidth]{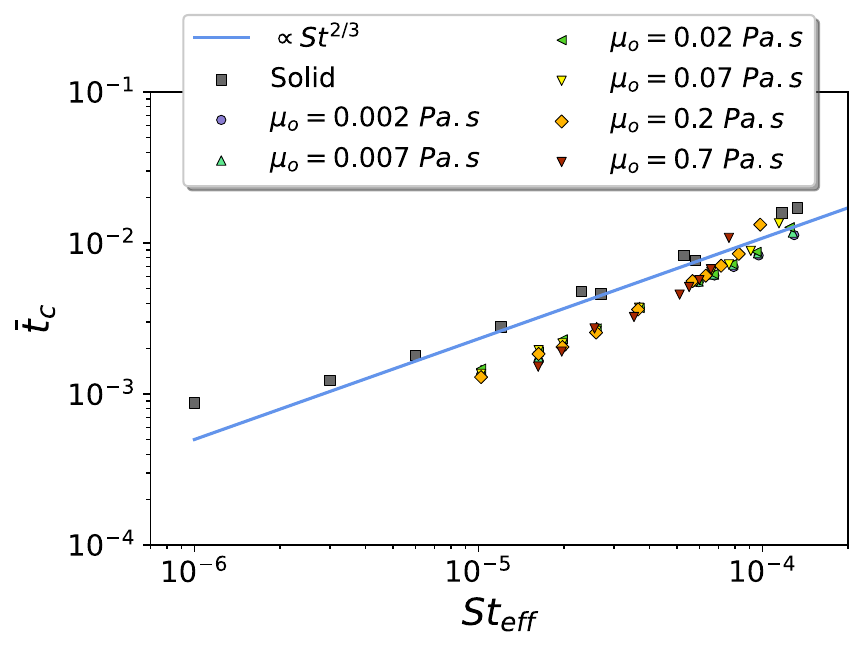} 
    \caption{(left) Raw data for the time of contact $t_c$ (right) Rescale impact time with the effective Stokes.}
    \label{figure 10}
\end{figure}

\begin{figure}
    \centering
    \includegraphics[width=0.45\linewidth]{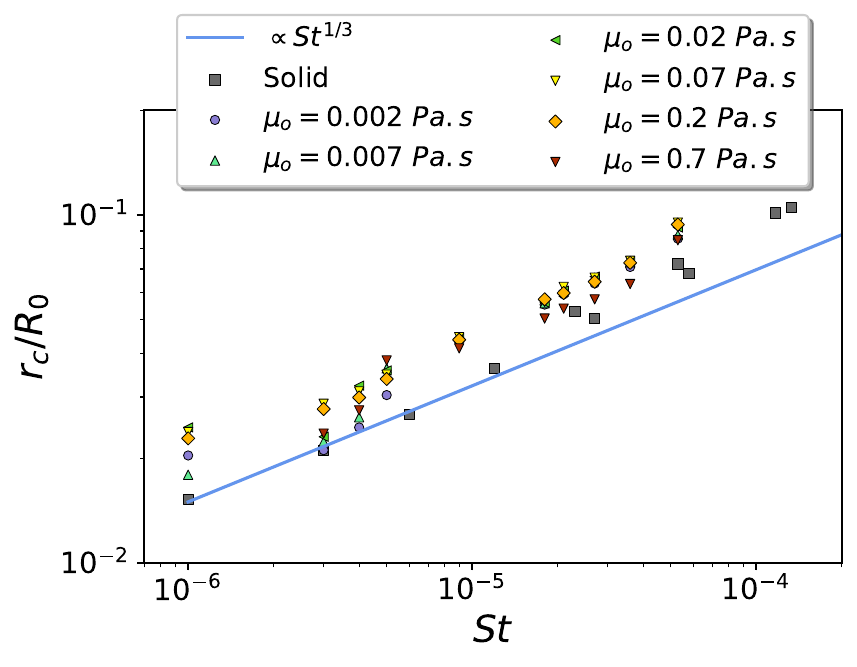}
    \includegraphics[width=0.45\linewidth]{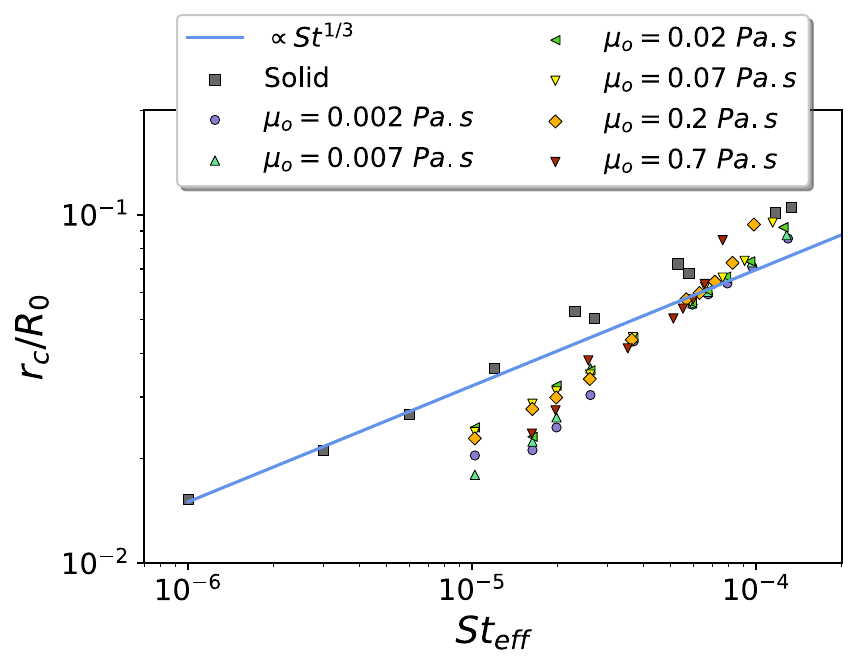} 
    \caption{(left) Raw data of the gas disc radius $r_c$ measure at the contact time $t_c$ . (right) Rescale radius with the effective Stokes}
    \label{figure 11}
\end{figure}

We present here in Figures \ref{figure 10} and \ref{figure 11} the rescaled results obtained for the time of the contact and the radius of gas disc. We clearly observe a good collapse of the data for the contact time $t_c$, which is in agreement with the measurement of $h_c$. For the radius, all the raw data aligns well with the $St^{1/3}$ and the solid surface data. However, the rescaling does not lead to a good collapse. This can be explained by the complex film deformation, since we are only considering the value of the deformation under the center of the droplet. In fact the film is deformed everywhere under the droplet, especially at the contact points, where a tiny dimple appears as shown in Figure \ref{fig : compa 2 visco}. Since this effect is not taken into account in our model, we cannot expect a good collapse of the data for the radius. Finally, although experimental measurements on the geometry of the entrapped bubble in such configuration do not exist yet, experimental comparisons should be available following the observations already done for same liquid impact on deep pool~\citep{tran_air_2013,thoraval_drop_2013}.

\newpage
\section{Conclusion}

We have investigated the influence of an immiscible thin film on the bubble entrapment dynamic following a droplet impact, focusing in particular on the role of the oil viscosity. We show that the lubrication of the gas layer is still the main effect controlling the dynamics of bubble formation, but that the oil viscosity plays an important role for understanding the film influence compared to the impact on a solid substrate. Indeed, the cushioning of the gas layer, when the drop approaches the oil film, does deform both the drop and the oil film. This second deformation is at the origin of the delay observed for the bubble entrapment. By accounting this deformation in the model, we obtain a new expression which fits well our data. 
Further work should investigate the influence of the other physical parameters, in particular the thickness of the oil layer \citep{hendrix_universal_2016} which should highlight the transition between the viscous and the inviscid regime for the oil film.
\section*{Declaration of Interests}
The authors report no conflict of interest.

\section*{Acknowledgement}
It is our pleasure to thank the Fonds Franco-Canadien pour la Recherche (FFCR) for its support through the project "Impact de gouttes sur film liquide".

\appendix
\counterwithin{figure}{section}
\section{}

\begin{figure}
    \centering
    \includegraphics[width=0.45\linewidth]{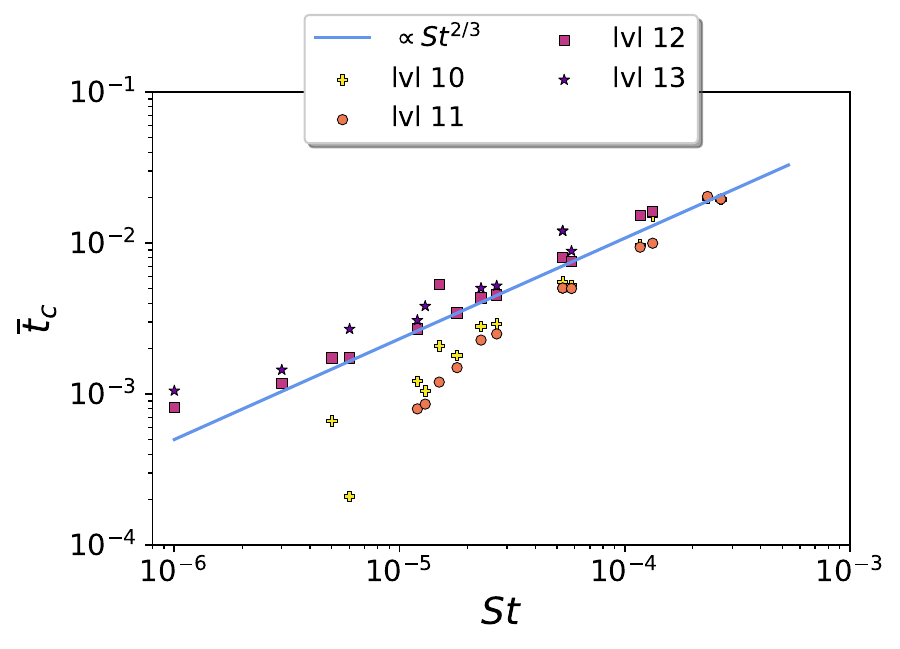} \includegraphics[width=0.45\linewidth]{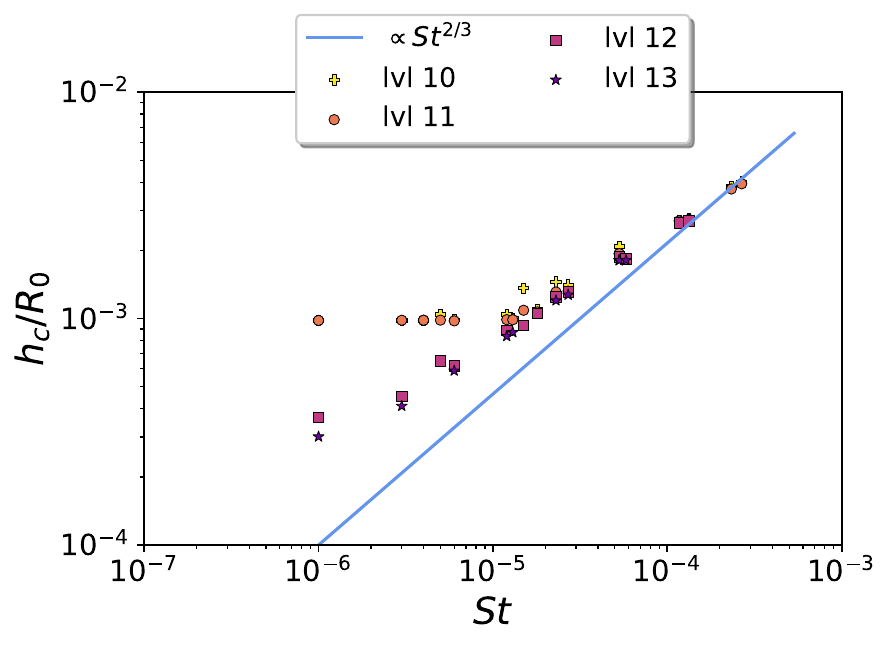} 
    \caption{(left) Dependence of the mesh refinement on the measurement of the contact : time $t_c$. (right) the thickness below the center of the droplet $h_c$ for a droplet impact over a solid substrate.}
    \label{figure 9}
\end{figure}

As explained in the main text, the contact is determined when the three phases are present in the same computational cell. We have performed a sensitivity study for the case of an impact over a solid surface to investigate the influence of the maximum mesh resolution. The contact is determined when the minimum thickness of the gas layer is below a threshold, basically equals to the size of the smallest cell. The Figures \ref{figure 9} shows the time of contact $t_c$, (left figure) and the central height of the gas layer $h_c$ (right figure) for different maximum mesh levels from levels 10 to 13. We observe a very low dependency on the mesh level for level 12 and above, justifying the use of level 12, in our computations.

\newpage
\bibliography{ref.bib}
\end{document}